\documentclass[%
  aps,
  prl,
  superscriptaddress, letterpaper,
  twocolumn,
  amsfonts, amssymb, amsmath]{revtex4-2}
  
\usepackage[utf8]{inputenc}
\usepackage{amsmath}
\usepackage{amsfonts}
\usepackage{amssymb}
\usepackage{graphicx,color}

\begin{document}

\title{From internal waves to turbulence in a stably stratified fluid}
\author{Costanza Rodda, Cl{\'e}ment Savaro, Vincent Bouillaut, Pierre Augier, Jo{\"e}l Sommeria, Thomas Valran, Samuel Viboud, and Nicolas Mordant}
\email[]{nicolas.mordant@univ-grenoble-alpes.fr}
\affiliation{Laboratoire des \'Ecoulements G\'eophysiques et Industriels, Universit\'e Grenoble Alpes, CNRS, Grenoble-INP,  F-38000 Grenoble, France}

\begin{abstract}
We report on the statistical analysis of stratified turbulence forced by large-scale waves. The setup mimics some features of the tidal forcing of turbulence in the ocean interior at submesoscales. Our experiments are performed in the large-scale Coriolis facility in Grenoble which is 13~m in diameter and 1~m deep. Four wavemakers excite large scale waves of moderate amplitude. In addition to weak internal wave turbulence at large scales, we observe strongly nonlinear waves, the breaking of which triggers intermittently strong turbulence at small scales. A transition to strongly nonlinear turbulence is observed at smaller scales. Our measurements are reminiscent of oceanic observations. Despite similarities with the empirical Garrett \& Munk spectrum that assumes weak wave turbulence, our observed energy spectra are rather be attributed to strongly nonlinear internal waves.
\end{abstract}


\maketitle

Understanding the energy balance in the ocean is a major challenge for efficiently modeling its dynamics for prediction purposes at various temporal scales, either connected to human life for seasonal forecast or to climate evolution scales \cite{mackinnon2017climate,wunsch2004vertical}. One issue is understanding and modeling small-scale dissipation processes of kinetic and potential energy. Oceanic submesoscales ($< 20$~km horizontal scales) are characterized by internal gravity waves that dissipate energy through nonlinear coupling processes (such as weak wave turbulence \cite{staquet2002internal,caillol2000kinetic,medvedev2007,nazarenko2011wave,polzin2011toward,dematteis2021downscale}) that induce fluxes of energy towards small scales at which molecular diffusion and viscosity absorb it. Internal waves can also become strongly nonlinear and break to generate directly small-scale strongly nonlinear turbulence. McKinnon et al. \cite{mackinnon2017climate} invoke many processes that contribute to the generation of internal waves. 
Linear internal waves have very specific properties \cite{Lighthill}, the first one being their dispersion relation $\omega=N\sin\theta$, where $\omega$ is the angular frequency, $\theta$ is the angle of the wave vector to the vertical and $N=\sqrt{-\dfrac{g}{\rho_0}\dfrac{\partial \rho}{\partial z}}$ is the Brunt-Vaisala frequency related to the vertical density gradient ($g$ is the acceleration of gravity, $\rho$ is the local density, $\rho_0$ is the reference density). The frequency is not related to the wavelength but only to the direction of propagation, which induces exotic features compared to simpler waves (acoustics or surface waves in a fluid, for example) such as anomalous reflection laws and linear attractors \cite{Maas2009,brouzet}. 

Laboratory experiments on gravity-dominated stratified turbulence are very challenging. They require to have both a large Reynolds number $Re=\frac{UL}{\nu}$ and a small Froude number $Fr=\frac{U}{NL}$, with $U$ an order of magnitude of the horizontal velocity, $L$ a horizontal length scale and $\nu$ the kinematic viscosity. This imposes in practice mild velocities ($\sim 1$~cm/s) and large scale experiments ($>1$~m) since $N$ is restricted for practical reasons to $N \lesssim 1$~rad/s and the viscosity of water is given $\nu\approx 10^{-6}$~m$^2$/s. These constraints explain why such flow regimes remain elusive in the laboratory. Here, we focus on the case of turbulence forced by waves in the spirit of oceanic turbulence forced by tide oscillations. Weak wave turbulence has been observed in the laboratory~\cite{davis2020succession,Lanchon2023,savaro2020generation,Rodda2022} or numerically \cite{LeReun}. In this study, we report experiments dedicated to strongly nonlinear stratified turbulence forced by large-scale waves. Previous experiments reported in \cite{savaro2020generation,Rodda2022} have shown that the large-scale motions of such a flow are made of weakly nonlinear wave turbulence. 
We focus on the strongest forcing amplitudes that can be reached in our setup. The values of the relevant dimensionless parameters ($Re$, $Fr$) are given in table~\ref{table_exps}. For all four experiments reported here, the Reynolds number is very large ($Re>10^4$) so that the flow is fully turbulent, and the Froude number is very low ($F_h\sim 10^{-2}$), which means that the dynamics is dominated by gravity and stratification. For stratified turbulence, the nonlinearity is better measured using the buoyancy Reynolds number $Re_b=ReF_h^2$ \cite{brethouwer2007}. In our experiments at the strongest forcing, $Re_b$ reaches values significantly larger than one, and hence the flow is expected to be in a strongly nonlinear regime of stratified turbulence. We report a multi-scale analysis of the flow, from several meters down to a few millimeters, to identify the nature of stratified turbulence forced by waves.

Experiments are performed in the Coriolis facility, a 13~m diameter, $H=1$~m deep tank in which density stratification can be set by using a vertically varying concentration of salt in water. Here, we use linear stratifications with associated Brunt-Vaisala frequencies $N=0.25$~rad/s. The setup is the same as that described in \cite{Rodda2022} and is only briefly recalled here (see also Supplemental Material (SM)). The flow is generated in a symmetric pentagonal subdomain with 6~m-long sides. Four vertical walls can oscillate around a horizontal axis at mid-depth of water to act as internal wave makers. The forcing parameters are the amplitude $\pm A$ of the top of the panel just above the water level and the center frequency $F$ of the forcing (normalized by $N$ so that $F$ is dimensionless). Available measurements are: (i) Particle Image Velocimetry (PIV) over 30 horizontal planes that fit in a square parallelepiped of size $3\times2.2\times 0.4$~m$^3$ approximately placed around the center of the pentagon (it provides the two horizontal components of velocity), (ii) Particle Image Velocimetry (PIV) in a vertical plane over a surface of horizontal length $0.54$~m and height $0.44$~m (it provides one horizontal component and the vertical component of the velocity) placed roughly at mid-depth of the water (iii) Vertical scans of the water density measured by two single point conductimeters of sensitive size 0.5~mm mounted on profilers that span the full depth of water (see SM for more details). Both PIV systems are undersampling the turbulent flow, but they are complementary as the vertical PIV is about 5 times better resolved than the horizontal one. Note that the PIV measurements are not simultaneous and were acquired in different experiments with identical parameters. Vertical PIV requires using water/alcohol/salt mixtures to provide optical index matching and prevent blurring of the images when imaging through the stratified water at large distances from the light sheet. The velocity of the vertical profiler of conductivity is 0.1 m/s, which is 10 times faster than the {\it rms} velocity of the fluid (at best 1~cm/s), so that the flow can be considered as frozen during the scan. The measurement of conductivity provides the local density $\rho$ of the fluid (which changes due to the addition of salt), from which buoyancy $b=-g\frac{\delta \rho}{\rho_0}$ can be calculated, where $\delta \rho=\rho-\overline{\rho}$ ($\overline{\rho}(z)$ is the average density profile).

\begin{table}[htb]
\begin{ruledtabular}
{\begin{tabular}{lcccccccccc}
 &$A$ (cm) & $$F$$& $Re$&$F_h$&$Re_b$ \\
\colrule
EXP1 & 9 & 0.39 &$4.3 \times 10^4$ & $7.4  \times 10^{-3}$ &2.4\\
EXP2& 5 & 0.67 & $1.9 \times 10^4$ & $1.5  \times 10^{-2}$ &4.4\\
EXP3 & 9 & 0.67 & $3.5 \times 10^4$ & $2.7  \times 10^{-2}$ &26\\
EXP4 & 7.5 & 0.67 & $2.9 \times 10^4$ & $2.3  \times 10^{-2}$ & 15
\end{tabular}}
\end{ruledtabular}
\caption{\label{table_exps}Parameters for the experimental runs. $A$ is the amplitude of oscillation of the wavemaker and $F$ is the normalized frequency of oscillation of the wavemaker. $Re=\frac{2ANH}{\nu}\sqrt{1-F^2}$ is the Reynolds number, $F_h=\frac{A}{2H}\frac{F^2}{\sqrt{1-F^2}}$ is the horizontal Froude number and $Re_b=ReF_h^2$ is the buoyancy Reynolds number \cite{Rodda2022}. These numbers have been defined using forcing parameters. $\nu$ is the kinematic viscosity. EXP4 is mostly a visualization experiment.}
\end{table}

\begin{figure}[!htb]
\includegraphics[width=8.5cm]{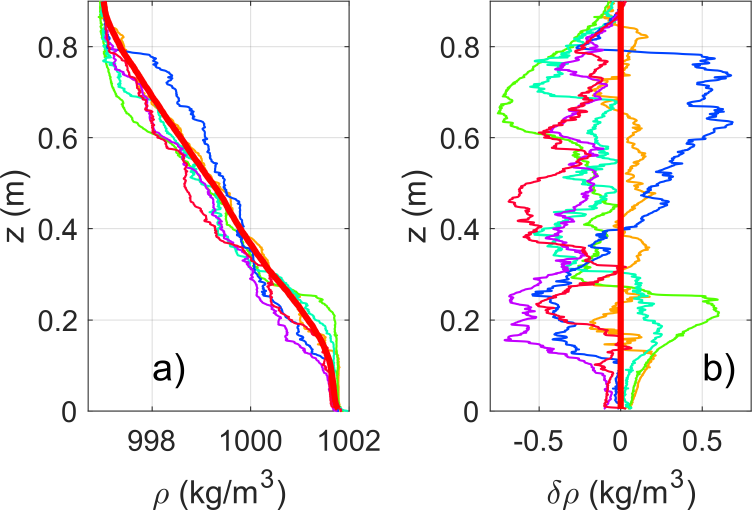}
\includegraphics[width=8.5cm]{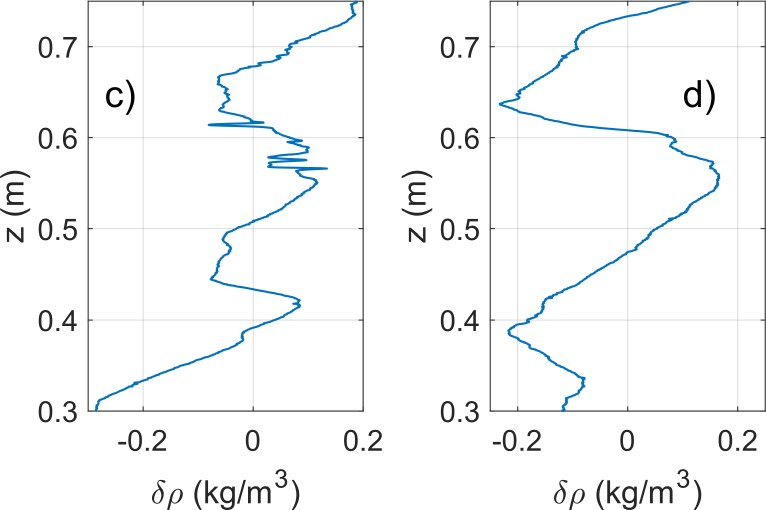}
\caption{\label{profiles}(a) Example of density profiles (EXP3) taken at various times. The thick red line is the average profile. It shows a mixed layer near the bottom due to the boundary layer. Statistics of buoyancy are computed for $0.3\,\textrm{m} <z<0.75\,\textrm{m}$. (b) same profiles after removing the average profile. (c) \& (d) are two examples of profiles of density fluctuations from EXP2 with a wave breaking phase between $z=0.55$~m and $0.6$~m (c) and without wave breaking (d).}
\end{figure}

\begin{figure}[!htb]
\includegraphics[width=8.7cm]{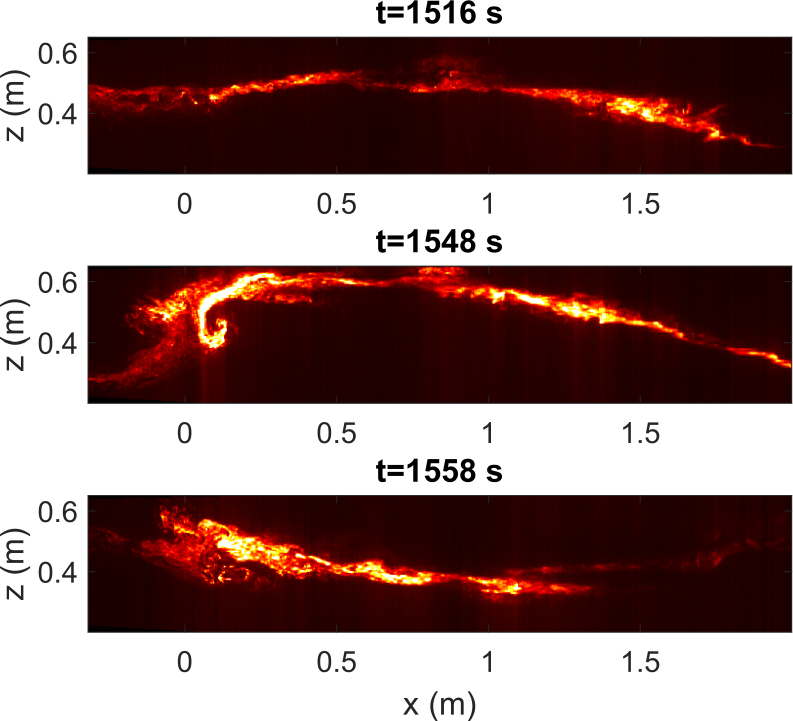}
\caption{\label{over}Snapshots in a vertical plane of a volumetric laser scan of a fluorescent dye layer (rhodamine) using the vertically scanning horizontal laser sheet in EXP4. A red filter has been put on the camera lens to remove the green light of the laser. See text for details. See SM for a movie.}
\end{figure}

\begin{figure}[!htb]
\includegraphics[width=8.5cm]{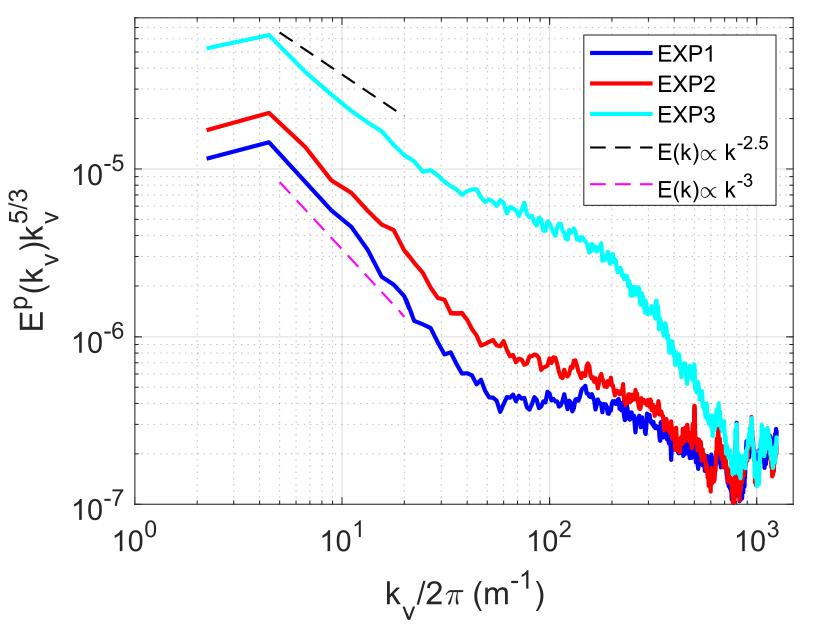}
\caption{\label{buospec}Vertical spectra of of the potential energy (for EXP1 to EXP3) computed as $\frac{1}{2} E^b(k_v)/N^2$ where $E^b(k_v)$ is the vertical buoyancy spectrum and $k_v$ is the vertical wavenumber.}
\end{figure}

Examples of density profiles are shown in figure~\ref{profiles}. 
At the largest $Re_b$ (EXP3 on the top row), the profiles of the density fluctuations appear to have a multiscale nature, with large-scale fluctuations due to the large-scale weakly nonlinear waves and small-scale fluctuations that point to the presence of very small-scale structures as well. For EXP2 (intermediate value of $Re_b$, bottom row), the scans show space-time intermittency. Some low activity periods exist, during which the scans are very smooth (d), while at some other times, the profiles show localized bursts of short-scale fluctuations separated by smoother regions (c). For EXP1 (lowest $Re_b$), the bursts are less frequent; for EXP3 (b), the short scales fluctuations seem always to be present. Figure~\ref{over} show a sequence of images of a layer of fluorescent rhodamine dispersed by a rather strong turbulence (EXP4). For visualisation purposes, a patch of rhodamine was injected at mid-depth of the tank with a density matching the one of the fluid at this depth. The patch of rhodamine is quickly dispersed horizontally by the turbulent motion, but due to the strongly anisotropic nature of stratified turbulence, it remains localized vertically for much longer. The image shows a vertical cut of a volumetric fluid scan using the same laser sheet as for horizontal PIV (but with 100 levels rather than 30). In the top image, the rhodamine layer remains localized in a very thin sheet that is undulating due to the large-scale internal waves. 32~s later, the layer is seen to roll over itself (left part of the image), highlighting a vortical structure that is due to an overturning internal wave. The bottom image shows the layer 10~s later than the middle one. One sees that the layer is much thicker in the region of the overturning wave due to the dispersion of the dye by the small-scale turbulence generated by the overturning waves. At later times (not shown), the dispersed dye forms again a thin layer similar to the top image due to the buoyancy restoring force that gathers the dispersed droplets of dye (see SM for a movie). Most likely, a part of the dye must have been mixed, but it is not visible in these images. Such events repeat at various times and positions. This visualization explains the space-time intermittency observed at intermediate values of $Re_b$ as the occurrence of internal wave breaking at various places in the flow. Figure~\ref{buospec} shows the vertical power spectrum of the density for EXP1 to EXP3 (compensated by $k_v^{5/3}$). The large-scale part displays a steep decay, possibly a power-law decay, with exponents $-3$ to $-2.5$. For EXP1 and EXP2, this decay is followed by a clear $k_v^{-5/3}$ range at centimeter scales and then an exponential-like fast decay due to the low pass filtering related to the time response of the probe during the scans. For EXP3, the $k_v^{-5/3}$ range is largely filtered out by the probe response, but a clear transition to a shallower decay is visible nonetheless. This $k^{-5/3}$ range indicates very strongly nonlinear turbulence of the Kolmogorov type that is expected at small scales in strongly nonlinear stratified turbulence \cite{brethouwer2007}.

\begin{figure}[!htb]
\includegraphics[width=8.5cm]{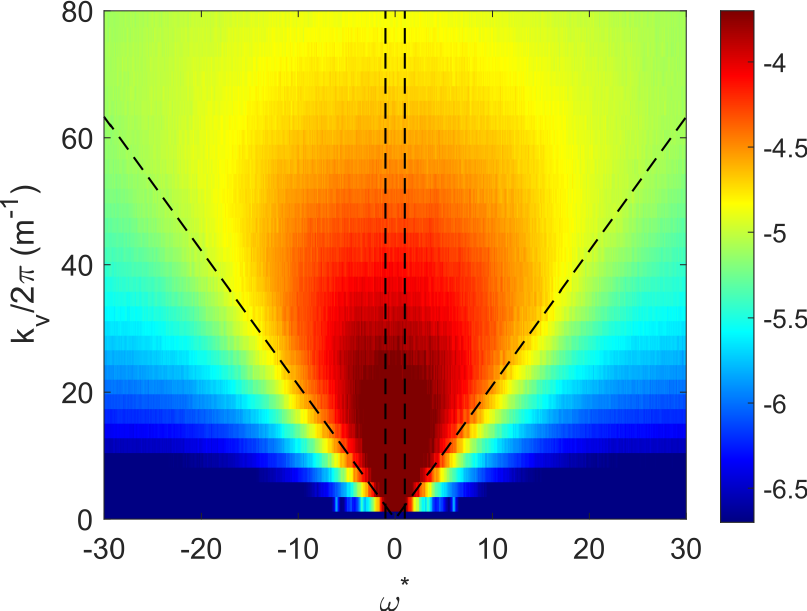}
\caption{\label{komv}$(k_v,\omega)$ spectrum of horizontal velocity $u$ from the vertical PIV data of EXP3. Vertical dashed lines mark $|\omega^*|=1$. Inclined dashed lines mark the limit $\omega=\sigma k$ where $\sigma=\sqrt{2u_{rms}^2+w_{rms}^2}$ is an estimate of the {\it rms} velocity magnitude. The spectrum has been compensated by $k_v^{-5/2}$. Colors are log$_{10}$-scale.}
\end{figure}

The impact of strong non-linearity is also visible in the velocity field. Figure~\ref{komv} shows the mixed $(k_v,\omega)$ spectrum of the horizontal velocity from the vertical PIV data of EXP3 ($k_v$ is the vertical wave number). For weakly nonlinear waves, the energy must remain confined at frequencies $-1<\omega^*<1$, which is the case at large scales $k_v\lesssim10\pi$ rad/m. At smaller scales (typically between 2 to 20 cm), energy is also present for frequencies $|\omega^*|>1$, which is a sign of a strong level of nonlinearity. Most of the energy lies in a wedge which is limited by $|\omega|\lesssim \sigma k_v$. This can be interpreted as a Doppler effect on the waves \cite{campagne2015disentangling}. In absence of waves a similar effect will be observed due to the sweeping of small-scale structures by the large-scale flow as is the case for regular Navier-Stokes turbulence without stratification. The latter case was shown in \cite{Clark2015} in numerical simulations, where such a wedge-shaped spectrum is visible as well despite the absence of waves. 

\begin{figure}[!htb]
\includegraphics[width=8.5cm]{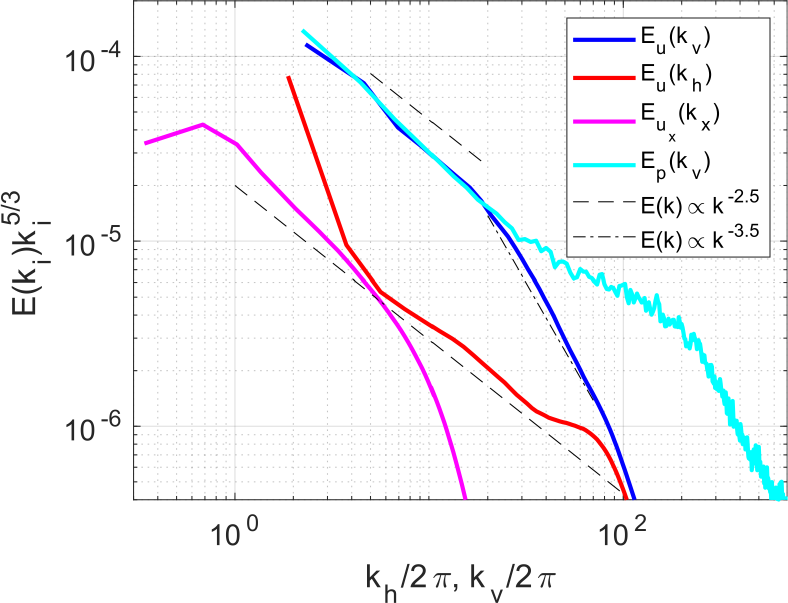}
\caption{\label{khkv}Spectra, either horizontal (function of $k_h$) or vertical (function of $k_v$) of one horizontal velocity component (for both PIV datasets) and of the potential energy (vertical spectrum only) for EXP3. Magenta: horizontal PIV data. Red \& blue: vertical PIV data. Cyan: density profiles.}
\end{figure}

Horizontal and vertical spectra of the kinetic and potential energy are shown in figure~\ref{khkv}. Horizontal spectra of the horizontal velocity are shown for both PIV measurements (magenta for large-scale horizontal PIV, red for vertical PIV). Both curves are restricted at high $k_h$ by the PIV resolution and by the occurrence of noise. The first points of the vertical PIV (red) are also most likely affected by the fact that very large scales exist but are not resolved due to the size of the image. Both curves are quite consistent and show a low $k_h$ ($k_h<10\pi$) decay compatible with a $k_h^{-2.5}$ decay. The transition to a plateau, i.e. a $k_h^{-5/3}$ decay, although expected from the phenomenology, may be rather due to experimental noise. The vertical spectrum of velocity also shows an initial decay compatible with $k_v^{-2.5}$ followed by a steeper decay. The two spectra show a very strong anisotropy at intermediate scales ($3<k/2\pi<20$) with a ratio between the two spectra close to a factor 10. Anisotropy decays at higher $k$. The potential energy spectrum shows a similar decay $k_v^{-2.5}$ in the range $10\pi<k_v<40\pi$ and is very close to the kinetic energy in that range, suggesting a range with almost equipartition of energy \cite{lindborg2006energy}. The small-scale spectrum of the flow is only accessible through the potential energy spectrum, and it shows the transition to a shallower spectrum, possibly $k_v^{-5/3}$. 


{\it Discussion.} Two frameworks coexist to discuss these results. On one side, oceanic submesoscale turbulent data are interpreted as internal wave turbulence and empirically described by the Garrett \& Munk (GM) spectrum \cite{garrett1979internal} that behaves as $E(k_v,\omega)\propto k_v^{-2}\omega^{-2}$ with some geographical or temporal dispersion of the values of the exponents~\cite{pollmann2020,leboyer2021}. On the other side, strongly non linear stratified turbulence (SNLST) has been described in \cite{brethouwer2007} with a horizontal spectrum of the horizontal velocity scaling as $k_h^{-5/3}$ and the vertical spectrum scaling as $k_v^{-3}$ followed by $k_v^{-5/3}$ at the smallest inertial scales at which turbulence becomes isotropic. Our experiment shows a $k_v^{-5/3}$ at the smallest scales associated with wave breaking that appears consistent with the second framework. However the horizontal spectrum remains steeper than $k_h^{-5/3}$ at the scales that we can resolve with our PIV. Although the decay of the horizontal spectrum is strongly evolving with $Re_b$ with an apparent exponent increasing from about $-4.5$ to $-2.5$ (see SM) we did not reach the SNLST scaling. 
{These observations are consistent with what discussed by \cite{Khani}, who showed how viscosity affects the spectra by steepening their slopes and introducing spurious energy cascades. Our values of $Re_b$ remain moderate and therefore viscous effects are likely to play a role in the horizontal dynamics and may explain a decay of the horizontal spectra that remains steeper than expected theoretically~\cite{bartello_tobias_2013,lindborg2006energy,Khani}. }
At the largest scales of fig.~\ref{khkv}, for EXP3, the exponent of the various spectra of horizontal velocity is close to $k_v^{-2.5}$, $k_h^{-2.5}$ (fig.~\ref{khkv}) or $\omega^{-2.5}$ (see SM), which exponent values are compatible with oceanic observations of GM spectrum (on the low side of the interval of observed exponents~\cite{pollmann2020,leboyer2021}). The vertical spectrum then transits to a steeper decay with a spectral exponent about $-3.5$. This scaling may be the precursor of the $k_v^{-3}$ scaling of the SNLST framework that can be also interpreted as a critical balance of strongly non linear wave turbulence \cite{nazarenko2011wave,nazarenko_critical_2011}. Our observations show that turbulence at these scales results indeed from strongly nonlinear internal waves (fig.~\ref{komv}). One may expect to reach a true $k_v^{-3}$ scaling at higher values of $Re_b$. Increasing further the value $Re_b$ in our experiment is challenging. The Froude number must remain small so we can not increase much further the amplitude of the forcing. We are left with increasing $N$ but this would require to use tremendous amounts of salt and alcohol. Even if the current value of $Re_b$ remains somewhat moderate we show a clear trend to the coexistence of a regime of weak turbulence at the largest horizontal scales, strongly non linear wave turbulence at intermediate scales followed by strongly non linear turbulence at the smallest scales generated through wave breaking. This global picture may be representative of the processes at play in the ocean interior that are believed to be of major importance for mixing of the oceanic stratification. Beyond the value of $Re_b$, our experiment is missing rotation which is also an important process in the ocean at low frequency. The Coriolis facility has been designed to rotate and adding rotation will be the object of future studies.

\begin{acknowledgments}
This project has received financial support from Fondation Simone et Cino Del Duca of the Institut de France, from the European Research Council (ERC) under the European Union's Horizon 2020 research and innovation program (grant agreement No 647018-WATU) and from the Simons Foundation through the Simons collaboration on wave turbulence. 
\end{acknowledgments}

\bibliography{biblio3}

\end{document}